\documentclass[aps,twocolumn,prl,amssymb,floatfix]{revtex4-1}

\usepackage{graphicx}

\begin{document}

\title{Fast electrochemical doping due to front instability in organic semiconductors}

\author{V. Bychkov$^{1}$, P. Matyba$^{1}$, V. Akkerman$^{2}$, M. Modestov$^{1}$, D.Valiev$^{1}$, G. Brodin$^{1}$, C. K. Law$^{2}$,
M. Marklund$^{1}$, and L. Edman$^{1}$}

\affiliation{$^{1}$Department of Physics, Ume{\aa} University,
SE-90187 Ume{\aa}, Sweden}

\affiliation{$^{2}$Department of Mechanical and Aerospace
Engineering, Princeton University, D323-A, Engineering Quad.,
Princeton NJ 08544-5263 USA}

\begin{abstract}
The electrochemical doping transformation in organic semiconductor devices is studied in application to light-emitting cells. It is shown that the device performance can be significantly improved by utilizing new fundamental properties of the doping process. We obtain an instability, which distorts the doping fronts and increases the doping rate considerably. We explain the physical mechanism of the instability, develop theory, provide experimental evidence, and perform numerical simulations. We further show how improved device design can amplify the instability thus leading to a much faster doping process and device kinetics.
\end{abstract}

\maketitle

\newpage

Organic semiconductors (OSCs) are expected to revolutionize everyday electronics by offering interesting and attractive properties which distinguish them from conventional inorganic semiconductors \cite{Pei-et-al-1995,Forrest-2004,Leger-2008}. In addition to simple processing, low cost, soft and conformable character, OSCs provide the intriguing possibility of \textit{in-situ} chemical and electrochemical doping. This doping transformation leads to significant changes in important material properties, such as conductivity, color, volume, and surface energy \cite{Forrest-2004,Leger-2008}. The opportunity for a controllable tuning of the properties of OSCs via doping has stimulated the emergence "organic electronics" with a large number of novel and flexible applications \cite{Pei-et-al-1995,Forrest-2004,Leger-2008,Arkipov-et-al-2005}. Operationally, electrochemical doping is performed by applying a potential to a metal electrode coated with an OSC in contact with an electrolyte. During n-type (p-type) doping, electrons (holes) are injected into the OSC from the cathode (anode) and subsequently compensated by redistribution of cations (anions) in the electrolyte \cite{Pei-et-al-1996,Gao-Dane-2004,Matyba-et-al-2008,Lin-et-al-2010,Matyba-et-al-2010}. When OSC is populated by light charges, its conductivity increases by two-three orders of magnitude in comparison with the original conductivity due to bulky ions. A complex manifestation of the electrochemical doping is the \textit{in-situ} formation of a dynamic p-n junction when p- and n-doping fronts meet in a light-emitting electrochemical cell (LEC) \cite{Rodovsky-et-al-2010,Matyba-et-al-2009,Slinker-et-al-2007,Costa-et-al-2010,Latini-et-al-2010,Inganas-2010,Hu-Xu-2010,Hoven-et-al-2010,Yu-et-al-2009}.
However, the electrochemical doping involves redistribution of heavy and slow ions, which makes such organic electronics devices slow to turn-on. The speed-up of the doping process is thus vitally important for technical applications. Here we obtain, theoretically and experimentally, instability of the doping transformation front, which leads to much faster kinetics of the process.  We suggest an improved device design on the basis of the attained fundamental insight. We show that a corrugated pattern in the electrode surface amplifies the instability and increases the doping rate; thus substantially reducing the device turn-on time.

The configuration of two doping fronts counter-propagating towards each other in an LEC device is schematically illustrated in Fig. 1(a). Fig. 1(b) shows the internal structure of two such planar fronts, as calculated using the method of \cite{Modestov-et-al-2010}. The doping fronts, however, can never be ideally planar because the unavoidable small perturbations grow with time and distort their shape. The physics behind this new instability is related to the phenomenon of St. Elmo's fire, caused by the increased electric field at a convex conducting surface. In the present configuration, any leading perturbation hump at a doping front causes local increase of the electric field in the undoped region. Since the front velocity is proportional to the electric field, see Eq. (1) below, this stronger electric field will force the front to propagate faster locally, thus producing a positive feedback and an unstable growth of the hump.

To study the instability growth, we employ the p/n-doping front velocity \cite{Modestov-et-al-2010}
\begin{equation}
\label{eq1}
{\rm {\bf U}}_{p,n} = \pm {\frac{{n_{0}}} {{n_{h,e}}} }(\mu _{ +}  + \mu _{
-}  ){\rm {\bf E}}_{0} ,
\end{equation}
where ${\rm {\bf E}}_{0} = - \nabla \phi _{0} $  is the electric field intensity just ahead of the front,  $\phi _{0} $ the electric field potential, $n_{0} $  the initial ion concentration in the pristine active material, $n_{h,e} $  the final concentration of the holes/electrons behind the front (determined by the thermodynamic properties of a particular OSC),  $\mu _{ +}  $ and $\mu _{ -}  $  the mobilities of the positive/negative ions. The "$\pm$" sign in Eq. (1) indicates that the p-type ($+$) and n-type ($-$) doping fronts propagate in opposite directions. We consider a perturbation Fourier mode  $\tilde {X}(t)\exp (iky)$ of a p-doping front $x = X_{p} (y,t)$  in Fig. 1(a), which induces also perturbations of the electrical field in the undoped region. The doped region may be treated as equipotential due to high conductivity. The front thickness may be characterized by a length scale  $L_{p} $, of about $10^{ - 4}{\rm -}  10^{ - 3} {\rm m}{\rm m}$, determined by ionic diffusion \cite{Modestov-et-al-2010}. Since the characteristic size of the experimentally observed front perturbations (approximately $0.2 \textrm{mm}$) is much larger than  $L_{p} $, then we may treat the front as infinitesimally thin. Most of the time the doping fronts are sufficiently far away from each other,  $k(X_{n} - X_{p})
\gg 1$, therefore, the instability of one front is not affected by the other. The linearized form of Eq. (1)
\begin{equation}
\label{eq2}
\partial _{t} \tilde {X} = - {\frac{{n_{0}}} {{n_{h}}} }(\mu _{ +}  + \mu
_{ -}  )\partial _{x} \tilde {\phi} ,
\quad
ikU_{p} \tilde {X} = {\frac{{n_{0}}} {{n_{h}}} }(\mu _{ +}  + \mu _{ -}
)\partial _{y} \tilde {\phi} .
\end{equation}
together with the solution to the Laplace equation for the electric potential in the udoped region, $\tilde {\phi}  \propto \exp (iky - {\left|
{k} \right|}x)$, yields the equation $\partial _{t} \tilde {X} = U_{p} {\left| {k} \right|}\tilde {X}$	 								
describing the perturbation growth with time. Thus, we obtain an instability for the p-type doping front, where the initially small perturbations grow exponentially as  $\tilde {X} \propto \exp(\sigma t)$, with the growth rate  $\sigma = U_{p} {\left| {k} \right|}$. The same result holds for the n-type doping front by replacing $U_{p}$  with  $U_{n}$. This new instability shows interesting mathematical similarities to the Darrieus-Landau instability encountered in combustion \cite{Zeldovich-et-al-1985,Bychkov-Liberman-2000} and inertial confined fusion \cite{Modestov-et-al-2009}.
\begin{figure}
\centering
\includegraphics[width=3.4in,height=3.8in]{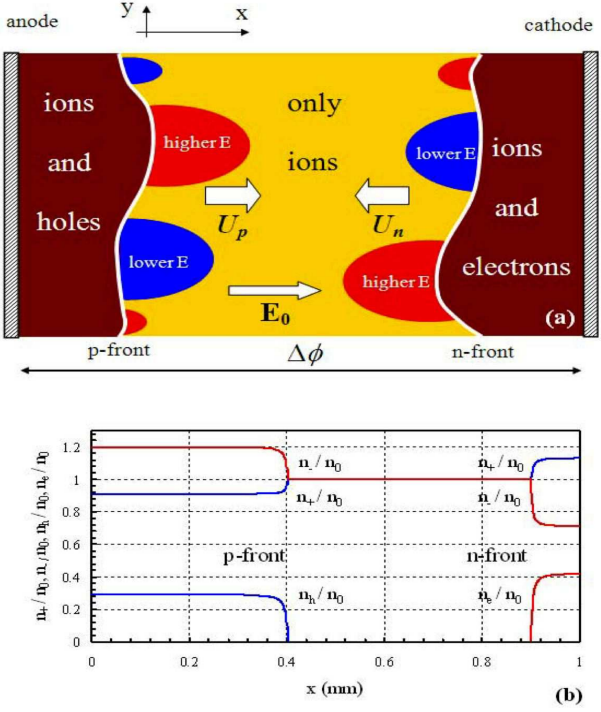}
\caption{Schematic of the p- and n-doping fronts in an
organic semiconductor film (a), and internal structure of the planar
fronts at 90 s (b).}
\end{figure}

To observe the instability experimentally, we utilize an open planar LEC device, comprising an   \{MEH-PPV + PEO + KCF$_{{\rm
3}}$SO$_{{\rm 3}}$\} active material positioned between two Au electrodes as described in \cite{Modestov-et-al-2010}. Two counter-propagating doping fronts can be distinguished in Fig. 2 (a, b); the dark regions with quenched fluorescence correspond to the doped  MEH-PPV; the electrodes are indicated with white dashed lines. When the fronts meet and form a p-n junction, recombination of subsequently injected holes and electrons can lead to light emission (see Fig. 2(c), taken in darkness). Figures 2 (a,b) clearly demonstrate that both doping fronts are unstable with respect to small perturbations, though with different outcomes at the nonlinear stage. The instability of the p-front produces smooth humps with a relatively large scale, about  $(0.1
- 0.{\rm 3}{\rm )} {\rm m}{\rm m}$, while the n-front appears as a combination of thin elongated "fingers". We focus on the p-front dynamics, since p-doping dominates for this choice of OSC sweeping more than 75\% of the active material \cite{Fang-et-al-2008}.

\begin{figure}
\centering
\includegraphics[width=3.4in,height=4.0in]{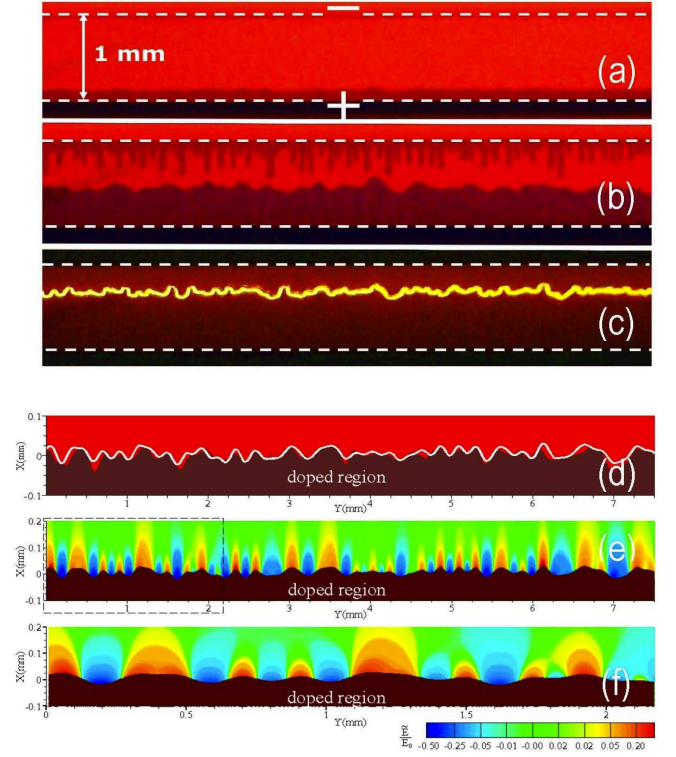}
\caption{Experimental photos of the doping fronts
demonstrate development of the instability at the initial stage (a),
$t = 46$ s, at the developed stage (b), $t= 80$ s, and the
dynamic p-n junction (c), $t > 170$ s. Plot (d) compares the
p-front shapes obtained in the simulations (white curve) and
experiments (shading) at 70 s. Plots (e), (f) show relative
increase of the electric field in the undoped region obtained
numerically for the whole front squeezed along Y-axis, (e), and for
the selected part with equal scales, (f).}
\end{figure}

A smooth shape of the p-doping front allows reducing the full model to a single nonlinear equation
\begin{equation}
\label{eq4}
U_{p}^{ - 1} \partial _{t} \tilde {X} = \hat {J}\tilde {X} + (\partial _{y}
\tilde {X})^{2} + U_{p}^{ - 1} \hat {H}{\left[ {\partial _{t} \tilde
{X}\;\partial _{y} \tilde {X}} \right]} + {\frac{{\lambda _{p}}} {{2\pi
}}}\partial _{yy}^{2} \tilde {X},
\end{equation}
where the Darrieus-Landau operator $\hat {J}$ and the Hilbert operator $\hat
{H}$ imply products by ${\left| {k} \right|}$ and ${\left| {k} \right|} / ik$
in Fourier space, respectively, and $\lambda _{p} \propto L_{p} $ is the
cut-off wavelength related to the small front thickness (detailed derivation will be presented elsewhere). Equation (\ref{eq4}) has been solved numerically with initial conditions extracted from the experimental data at $t = 10$ s. The numerical modeling reproduces well  the characteristic shape of the p-front at later instants, e.g. at $t = 70$ s, see Fig. 2(d). Figs. 2(e) and (f) show the computed electric field ahead of the corrugated p-type front. The observed increase in the electric field at the humps of the front demonstrates the instability mechanism, as explained earlier. The analytical, experimental, and numerical results for the doping front positions are presented in Fig. 3. The analytical result (solid lines) shows two planar fronts accelerating towards each other. The acceleration takes place because a constant potential difference is effectively applied over the continuously decreasing distance between the doping fronts \cite{Robinson-et-al-2006,Smela-2008}, and because the front velocities are controlled by the potential gradient, Eq. (1). The analytical result relies on our experimental data \cite{Modestov-et-al-2010}, with the average hole concentration  $n_{h} = 8.6 \cdot 10^{25}m^{ - 3}$.
\begin{figure}
\centering
\includegraphics[width=3.4in,height=2.4in]{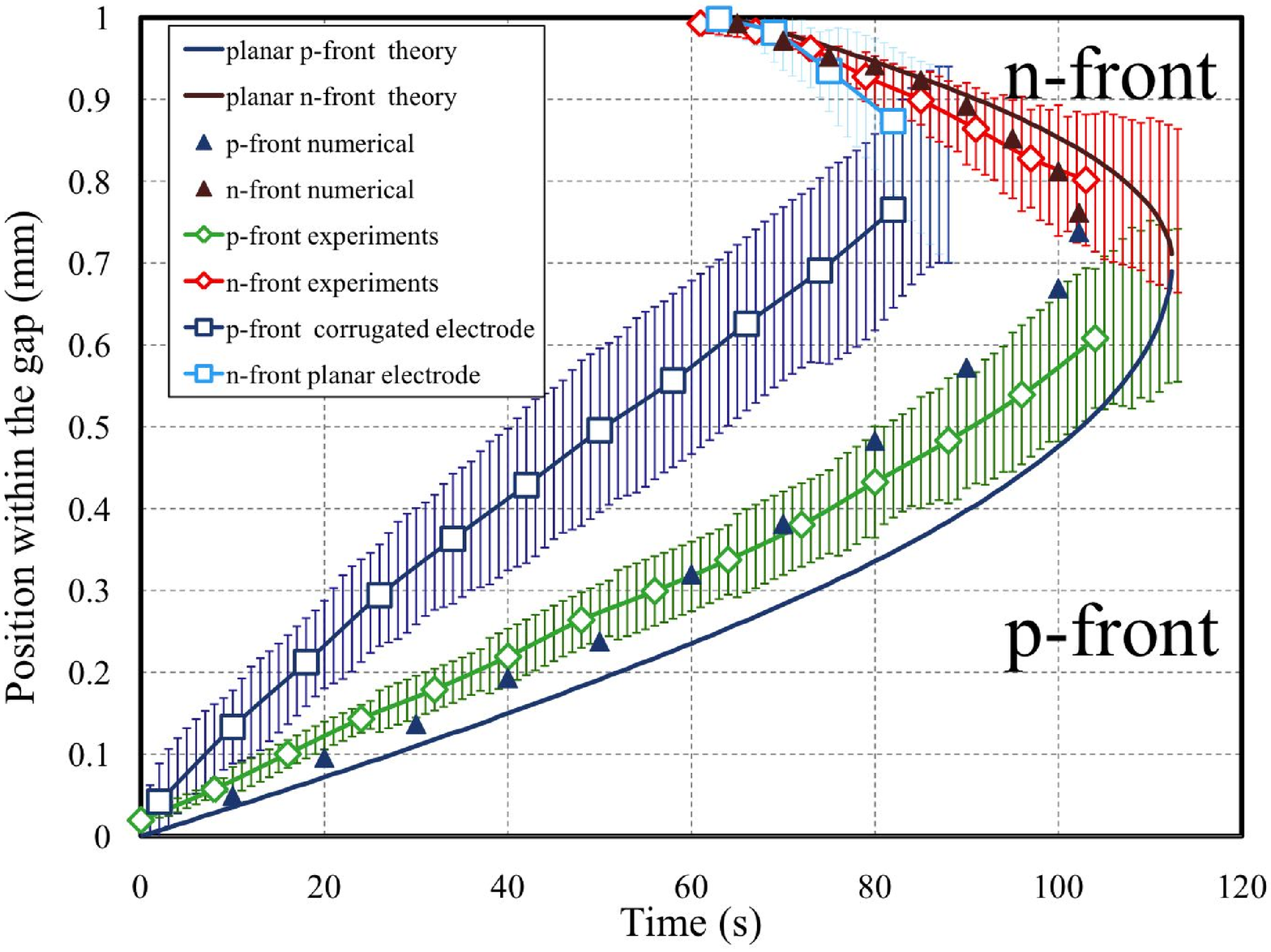}
\caption{Mean front position for p- and n-fronts versus time
obtained experimentally; predicted
theoretically for the planar fronts;
and calculated numerically for the wrinkled fronts (see the legend). The error bars of the experimental data indicate
difference in the positions of the fastest and slowest parts of the
corrugated front brush.}
\end{figure}

The experiments reveal a noticeably faster propagation of the non-planar doping fronts compared to the analytical result for planar fronts. In Fig. 3, open diamonds show the mean positions of the experimental non-planar doping fronts, where the "uncertainty" bars indicate the difference between the fastest and the slowest parts of the front brush. The difference can be attributed to the developing instability and the associated wrinkled front, which is not accounted for in the analytical description of the planar fronts. The analytical prediction for positions of the planar fronts correlates well with the slowest points of the experimentally observed fronts, while the leading points in experiments move approximately 1.4 times faster. In contrast, the numerical modeling does account for the instability and shows good agreement with the experiments.
\begin{figure}
\centering
\includegraphics[width=3.4in,height=3.4in]{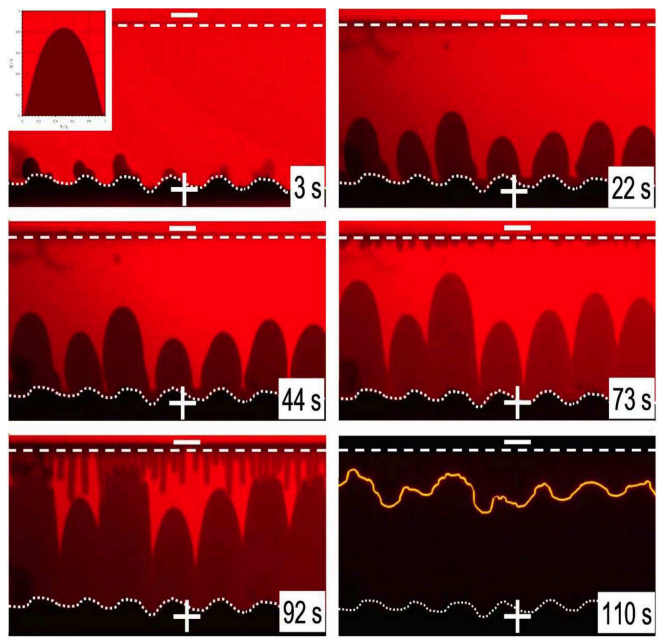}
\caption{Experimental photographs showing the time evolution
of the p-type doping front in an LEC device with the initial
corrugations of the anode. The corresponding time instances are
indicated on the photographs. The insert shows characteristic shape
of a strongly curved hump obtained in the theory.}
\end{figure}

The present results therefore verify that front instability increases the doping rate. However, the instability theory also indicates that the time and space available for the perturbation growth in the employed LEC devices are not sufficient to achieve a strongly curved front. Specifically, the theory suggests that the instability may double the velocity of a curved front in comparison to a planar one, which is not observed in the experimental results of Fig. 3 discussed above. Such strong increase of the front velocity happens due to strongly curved humps, which may develop at the front (see the insert of Fig. 4).
Therefore, to achieve a strongly curved front and a faster turn-on, we have triggered the instability growth by modifying the initial conditions so that the slowest initial stage of hump growth from natural "white-noise" perturbations is eliminated. Figure 4 presents results from an experiment with a corrugated pattern introduced as the surface of the anode (see the dotted line); the corrugation size was comparable to the humps observed in Fig. 2. The corrugations, though negligible in comparison with the total (1 mm) width of the active material, were anticipated to give rise to the desired initial perturbations of the p-type front. Confirming our anticipation, we observed strongly curved and rapidly growing humps at the p-type front, see Fig. 4. The position of the humps in this experiment is also included in Fig. 3, which shows that, in devices with a corrugated electrode surface, the doping rate may increase significantly; by a factor of two in the present case. We have also performed experiments with corrugations included into both the cathode and anode, and, as expected, we observed intensified front dynamics and much faster turn on. By introducing corrugations on the cathode we also find that the initial delay in the development of the n-type front was reduced.

To summarize, we have demonstrated theoretically and experimentally that electrochemical doping fronts are unstable. The obtained instability distorts the fronts and increases the doping rate considerably. This understanding of new fundamental properties of organic semiconductors was utilized in the design of light-emitting electrochemical cells with distinctly improved turn-on times. We further propose that similar design principles can be employed in other kinetically limited electrochemical devices, such as actuators \cite{Yumusak-Sariciftci-2010} and transistors \cite{Edman-et-al-2004,Bolin-et-al-2009,Larsson-et-al-2009,Kaihovirta-et-al-2010}.

The authors are grateful to Kempestiftelserna,
Carl Tryggers Stiftelse, and the Swedish Research Council (VR)
for financial support. L.E. is a ``Royal Swedish Academy of Sciences
Research Fellow'' supported by a grant from the Knut and Alice Wallenberg
Foundation. The work at Princeton University was supported by the Combustion
Energy Frontier Research Center funded by the Office of Basic Energy
Sciences of the U.S. Department of Energy.

\end{document}